\documentclass[aps,english,floatfix,amsmath,superscriptaddress,tightenlines,twocolumn,nofootinbib]{revtex4-2}
\usepackage{CJK} 
\usepackage{amssymb,amsthm,tikz,mathtools} 
\usetikzlibrary{decorations.markings,decorations.pathmorphing}
 \allowdisplaybreaks 
\usepackage[utf8]{inputenc}
\usepackage{hyperref}
\hypersetup{colorlinks=true,linkcolor=red,citecolor=magenta,urlcolor=blue}
\usepackage{url} 
\usepackage{xcolor} 
\usepackage{cleveref}
\usepackage{float} 
\usepackage[caption=false]{subfig}
\usepackage{makecell}
\usepackage{booktabs, multirow} 

\newtheorem{remark}{Remark}

\newcommand{\cH}{\mathcal{H}}
\newcommand{\cX}{\mathcal{X}}
\newcommand{\cZ}{\mathcal{Z}}
\newcommand{\ZZ}{\mathbb{Z}}
\DeclareMathOperator{\Tr}{Tr}
\newcommand{\<}{\langle}
\renewcommand{\>}{\rangle}
\newcommand{\const}{\textnormal{const.}}

\def\wtK{\widetilde K}
\newcommand{\ket}[1]{|#1\rangle}

\definecolor{BSorange}{RGB}{140,50,0}

\newcommand{\version}[1]{#1}

\def\ii{{\bf i}}

\begin{document}

\begin{CJK*}{UTF8}{bsmi}
\title{Conformal Field Theory Ground States as Critical Points of an Entropy Function}
\author{Ting-Chun Lin（林鼎鈞）}
\affiliation{Department of Physics, University of California at San Diego, La Jolla, CA 92093, USA}
\affiliation{Hon Hai Research Institute, Taipei, Taiwan}
\author{John McGreevy}
\affiliation{Department of Physics, University of California at San Diego, La Jolla, CA 92093, USA}
\date{\today}


\begin{abstract} 
  We derive an entropy formula satisfied by the ground states of 1+1D conformal field theories.
  The formula implies that the ground state is the critical point of an entropy function.
  We conjecture that this formula may serve as an information-theoretic criterion for conformal field theories, which differs from the conventional algebraic definition.
  In addition to these findings, we use the same proof method to extract the six global conformal generators of the conformal field theory from its ground state.
  We validate our results by testing them on different critical lattice models with excellent agreement.
\end{abstract}

\maketitle
\end{CJK*}

Quantum entanglement and quantum information have played important roles in the study of quantum matter.
For 2+1D gapped phases, this includes topological entanglement entropy \cite{kitaev2006topological,levin2006detecting}, entanglement spectrum \cite{li2008entanglement}, and the recent work on chiral central charge \cite{kim2022chiral} and minimal total central charge \cite{Siva:2021cgo}.
For 1+1D conformal field theories (CFTs), the entanglement entropy \cite{holzhey1994geometric,calabrese2004entanglement,calabrese2009entanglement} of the ground state is related to the central charge.
These tools are useful to distinguish quantum phases analytically and numerically.

More ambitiously, one may wonder if the reverse holds, namely, could there be entanglement conditions that are satisfied and only satisfied by ground states for certain quantum phases?
One proposal is given by Shi, Kato, and Kim \cite{shi2020fusion} where they stated two conditions that are conjectured to be satisfied and only satisfied by the ground states of topological quantum field theories\footnote{More precisely, any state that satisfies the conditions is the ground state of a Hamiltonian that belongs to a topological phase. And any topological phase has a Hamiltonian whose ground state satisfies the conditions.}.
Remarkably, using the two conditions, they are able to derive many known properties of 2+1D topological orders.
Another nice feature is that because the two conditions only involve the entropies of local regions, the conditions can be checked easily.
This program is called entanglement bootstrap and can be applied to other settings, including gapped domain walls \cite{shi2021entanglement} and higher dimensions \cite{huang2021knots}. 

Given the success of entanglement bootstrap for gapped topological orders,
  one may wonder if a similar set of entropy conditions exists for gapless states.
In this work, we propose a new set of ultra-violet (UV)-independent entropy conditions that apply to the ground states of 1+1D unitary CFTs.
The conditions additionally hold for 1+1D gapped phases at RG fixed points.
In the spirit of entanglement bootstrap,
we conjecture that these conditions characterize the ground states of 1+1D RG fixed points with Lorentz symmetry.\footnote{\version{Notice that in this statement we make a distinction between CFT and RG fixed point.
In fact we know examples of lattice wavefunctions that satisfy \Cref{eq:main-eq} that are not CFT ground states.
  Nevertheless, they are renormalization group fixed points.

To be clear, for the more specific case of CFT, this paper shows that \Cref{eq:main-eq} is a necessary condition,
  and we conjecture that \Cref{eq:main-eq} can also be viewed as a sufficient condition for RG fixed points.
Furthermore, the following argument suggests that
demanding \Cref{eq:main-eq} for multiple choices of regions may be a sufficient condition for CFT.

The condition \Cref{eq:main-eq} is a strong constraint;
 the fact that there are solutions to \Cref{eq:main-eq} is itself surprising.
Notice that each vector equation has roughly the same number of constraints as the number of variables in $|\psi\>$, and therefore imposing the vector equation for multiple choices of intervals is overconstrained.
When the state has $\ge 6$ intervals, if we require the vector equation for {\it every} choice of three contiguous intervals,
  there are at least $2$ independent vector equations, generically.
That means, in general, we do not expect any $|\psi\>$ that satisfy every vector equation to exist.
  However, as we have shown, each CFT ground state is a solution to the vector equations.
This is suggesting that these vector equations have intricate relations that are yet to be understood,
  and it is possible that this relation between equations has something to do with the Virasoro algebra.
}}

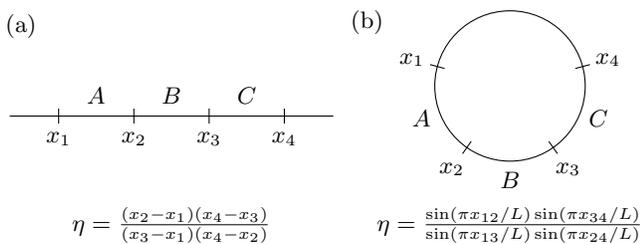
\begin{figure}
  \centering
  \begin{tikzpicture}[scale=0.5]
    \draw (-4,2.3)node{(a)};
    \begin{scope}[shift={(0,-0.1)}]
      \draw (-4.3,0) -- (4.3,0);
      \draw (-3,-0.2) node[anchor=north]{$x_1$} -- (-3,0.2)
            (-1,-0.2) node[anchor=north]{$x_2$} -- (-1,0.2)
            (1,-0.2) node[anchor=north]{$x_3$} -- (1,0.2)
            (3,-0.2) node[anchor=north]{$x_4$} -- (3,0.2);
      \draw (-2,0.1)node[anchor=south]{$A$}
            (0,0.1)node[anchor=south]{$B$}
            (2,0.1)node[anchor=south]{$C$};
    \end{scope}
    \begin{scope}
      \draw (0,-3)node{$\eta = \frac{(x_2-x_1)(x_4-x_3)}{(x_3-x_1)(x_4-x_2)}$};
    \end{scope}
    \draw (5.2,2.4)node{(b)};
    \begin{scope}[shift={(9,0.7)}]
      \draw (0,0) circle (2cm);
      \draw (165:1.8) -- (165:2.2)
            (235:1.8) -- (235:2.2)
            (305:1.8) -- (305:2.2)
            (15:1.8) -- (15:2.2);
      \draw (165:2.7)node{$x_1$}
            (235:2.7)node{$x_2$}
            (305:2.7)node{$x_3$}
            (15:2.7)node{$x_4$};
      \draw (200:2.5)node{$A$}
            (270:2.5)node{$B$}
            (340:2.5)node{$C$};
    \end{scope}
    \begin{scope}[shift={(9,0)}]
      \draw (0,-3)node{$\eta = \frac{\sin(\pi x_{12} / L)\sin(\pi x_{34} / L)}{\sin(\pi x_{13} / L)\sin(\pi x_{24} / L)}$};
    \end{scope}
  \end{tikzpicture}
  \caption{(a) Three consecutive intervals on an infinite system with the corresponding cross ratio $\eta$. (b) Three consecutive intervals on a circle with circumference $L$. Here, $x_{ij} \coloneqq x_j - x_i$ denotes the distance between $x_i$ and $x_j$.}
  \label{fig:3-intervals}
\end{figure}

Let $A, B, C$ be three consecutive intervals; see \Cref{fig:3-intervals}.
Our essential observation is that the ground state of a 1+1D unitary CFT $|\psi\>$ satisfies
\begin{equation} \label{eq:main-eq}
  K_\Delta \propto I \qquad\text{and}\qquad K_\Delta |\psi\> \propto |\psi\>\footnote{\version{Even though the operator equation $K_\Delta \propto I$ implies the vector equation $K_\Delta |\psi\> \propto |\psi\>$, we find the vector equation to be more robust than the operator equation. Therefore, we write both.}}\footnote{\version{In \cite{supplemental}, we give an argument that Eq.~\eqref{eq:main-eq} for different choices of $A,B,C$ exhausts the linear relations amongst single-interval entanglement Hamiltonians in a CFT groundstate.}}
\end{equation} 
where
\begin{align}
  K_\Delta
  & \coloneqq (K_{AB} + K_{BC}) - \eta (K_A + K_C) \nonumber\\
  & \qquad - (1-\eta) (K_B + K_{ABC})
\end{align}
and $K_X \coloneqq - \log \rho_X$\footnote{The reader might worry that $- \log \rho_X$ is not well-defined when $\rho_X$ has zero eigenvalues.
  Here, we note that the equation $K_\Delta |\psi\> \propto |\psi\>$ remains well-defined, because it projects out the problematic eigenvectors.
  This point is discussed in further detail after the derivation of the main results.}
is the entanglement Hamiltonian of the reduced density matrix $\rho_X \coloneqq \Tr_{\bar X} |\psi\>\<\psi|$,
$I$ is the identity operator,
and $\eta$ is the cross ratio of the intervals.
Moreover, when the central charge $c$ is known, the proportionality constant is given by
\begin{equation} \label{eq:main-eq-with-ratio}
  K_\Delta = \frac{c}{3} h(\eta) \qquad\text{and}\qquad K_\Delta |\psi\> = \frac{c}{3} h(\eta) |\psi\>
\end{equation}
where $h(\eta) \coloneqq - \eta \log \eta - (1-\eta) \log (1-\eta)$ is the binary entropy function.
Finally, we observe that Eq.~\eqref{eq:main-eq} is the condition for $|\psi\>$ to be a critical point of the following function:
\begin{align}
  S_\Delta(|\psi\>)
  & \coloneqq (S_{AB} + S_{BC}) - \eta (S_A + S_C) \nonumber\\
  & \qquad - (1-\eta) (S_B + S_{ABC}) \label{eq:saddle-point}
\end{align}
where $S_X \coloneqq S(\rho_X)$ is the von Neumann entanglement entropy between $X$ and $\bar X$.
The function is nonnegative because it is a convex combination of two nonnegative quantities, by weak monotonicity $S_{AB} + S_{BC} - S_A - S_C \ge 0$ and strong subadditivity $S_{AB} + S_{BC} - S_B - S_{ABC} \ge 0$ \cite{lieb1973proof,nielsen2002quantum}.
All the statements above can be extended to the ground state on a finite circle and the thermal state on an infinite line.

The formulae discussed above were derived in continuum CFT.  In many physical realizations, CFT arises as an approximation to a lattice model.  Importantly,
we show numerically that the equations hold approximately for various lattice models and the error $\Big\lvert K_\Delta |\psi\> - \frac{c}{3} h(\eta)|\psi\> \Big\rvert$ decays as the number of sites increases (as a power law).

Therefore, after obtaining the basic formulas, we explore their implications on states supported on a discrete lattice, by the following device: divide up the spatial direction into a collection of line segments, and regard each segment as a site of a lattice model with infinite-dimensional local Hilbert space.
We thereby identify new equations that could potentially hold for lattice models with finite-dimensional local Hilbert space, which can then be validated numerically.
Specifically, we express the entanglement Hamiltonians of intervals and the six global conformal generators as a combination of 1-site and 2-site entanglement Hamiltonians of the CFT ground state.
This has two important implications.
First, when the state satisfies the entropy formula, the six global conformal generators have the expected scaling properties under real-space RG.
This provides evidence that a state satisfying the entropy formula enjoys conformal symmetry.
Second, the result yields an algorithm that can reconstruct the CFT local Hamiltonian from the CFT ground state.
This is remarkable since the parent Hamiltonian construction \cite{perez2006matrix} that applies to matrix product states cannot be applied to gapless phases due to their long-range correlations.
This algorithm offers an alternative method for recovering the local Hamiltonians from states near RG fixed points, including gapped and gapless phases.  This result corroborates the general principle that a single representative wavefunction contains all the universal data about a state of matter.


We want to highlight an implicit theme of this work, which aims to go beyond the traditional algebraic formulation of CFT and provide an alternative analytic formulation.
Typically, CFTs are defined using algebras with equalities, which do not allow for the discussion of \emph{approximate CFTs}.
However, many examples exist that we would like to categorize as approximate CFTs, such as QFTs that are slight perturbations of CFTs or critical lattice models that are CFTs in the IR limit.
Because these models have different Hilbert spaces, finding a criterion that applies to all cases is challenging.
To address this issue, we utilize the entanglement entropy and entanglement Hamiltonian, which are agnostic to the Hilbert space.
We propose to define approximate CFTs as systems whose ground state approximately satisfies Eq.~\ref{eq:main-eq}.
Further research is needed to determine the usefulness of this proposal.


\section{CFTs and their entanglement properties}
Conformal field theories (CFTs) are field theories with scale invariance.
Such scale invariance often implies a larger invariance called the conformal symmetries, which include transformations that locally look like rescalings.
These conformal symmetries appear naturally in many physical systems, including the fixed points of the RG flow and the critical points of statistical models.

For 1+1D CFT, the entanglement entropy \cite{calabrese2009entanglement} and the entanglement Hamiltonian (EH) \cite{cardy2016entanglement} of the ground state on an interval $[x_1, x_2]$ are known to be
\begin{equation} \label{eq:entropy}
  S_{[x_1,x_2]} = \frac{c}{3} \log \frac{x_2-x_1}{\epsilon}
\end{equation}
and
\begin{equation} \label{eq:EH} 
  K_{[x_1,x_2]} = 2 \pi \int_{x_1}^{x_2} dx \frac{(x-x_1) (x_2-x)}{x_2-x_1} T_{00}(x) + \const
\end{equation}
where $c$ is the central charge, $\epsilon$ is the uniform UV cutoff, $T$ is the stress-energy tensor, and $\const$ is a number that depends on the UV cutoff $\epsilon$.
These two equations are the key to showing the main results.
As we will see, even though both equations suffer from UV divergences, the combinations in Eq.~\eqref{eq:main-eq} and Eq.~\eqref{eq:main-eq-with-ratio} are free from UV divergences.


\section{Derivation of the main results}
We first show Eq.~\eqref{eq:main-eq} and Eq.~\eqref{eq:main-eq-with-ratio} for the ground state of a 1+1D CFT on an infinite system. Then we show the equivalence of Eq.~\eqref{eq:main-eq} to the critical point condition of $S_\Delta$ defined in Eq.~\eqref{eq:saddle-point}.
Let $A = [x_1, x_2]$, $B = [x_2, x_3]$, and $C = [x_3, x_4]$ be three consecutive intervals.
For convenience, we define the function $f_{[x', x'']}(x) = \frac{(x-x')(x''-x)}{x''-x'} 1_{[x', x'']}$ where $1_{[x', x'']}$ is the indicator function.
By Eq.~\eqref{eq:EH},
\begin{align}
  &(K_{AB} + K_{BC}) - \eta (K_A + K_C) - (1-\eta) (K_B + K_{ABC}) \nonumber\\
  &\hspace*{1.5em} = \int_{-\infty}^{\infty} dx\, \big((f_{AB} + f_{BC}) - \eta (f_A + f_C) \nonumber\\
  &\hspace*{5em} - (1-\eta) (f_B + f_{ABC})\big)T_{00}(x) + \const
\end{align}
A straightforward calculation shows $(f_{AB} + f_{BC}) - \eta (f_A + f_C) - (1-\eta) (f_B + f_{ABC}) = 0$ from which \Cref{eq:main-eq} follows.

To obtain the ratio in Eq.~\eqref{eq:main-eq-with-ratio}, we multiply $\<\psi|$ on both sides.
Because $\<\psi|K_{A}|\psi\> = S_{A}$, the left hand side becomes
$(S_{AB} + S_{BC}) - \eta (S_A + S_C) - (1-\eta) (S_B + S_{ABC})$
which evaluates to $\frac{c}{3} h(\eta)$ using Eq.~\eqref{eq:entropy}.

Now we show that Eq.~\eqref{eq:main-eq} is precisely the condition
for vanishing variation of $S_\Delta$ with respect to the state.
Recall that $|\psi\>$ is a critical point if the gradient is $0$ subject to $(\<\psi| + \<d \psi|)(|\psi\> + |d\psi\>) = 1$, i.e., $\<d \psi|\psi\> + \<\psi|d\psi\> = 0$.
To compute the gradient of $S_\Delta = (S_{AB} + S_{BC}) - \eta (S_A + S_C) - (1-\eta) (S_B + S_{ABC})$ we use the first order derivative of the entanglement entropy:
\begin{equation} \label{eq:entropy-first-order}
  d S_X(|\psi\>) = \<d \psi|K_X|\psi\> + \<\psi|K_X|d \psi\>
\end{equation}
as reviewed in the Supplementary Material.
Therefore, the critical point $|\psi\>$ has
\begin{equation}
  d S_\Delta = \<d \psi|K_\Delta|\psi\> + \<\psi|K_\Delta|d \psi\> = 0
\end{equation}
for all $|d \psi\>$ satisfying $\<d \psi|\psi\> + \<\psi|d\psi\> = 0$.
Recall $K_\Delta = (K_{AB} + K_{BC}) - \eta (K_A + K_C) - (1-\eta) (K_B + K_{ABC})$.
This is equivalent to $K_\Delta |\psi\> \propto |\psi\>$ in Eq.~\eqref{eq:main-eq}.
Note that the left equation in Eq.~\eqref{eq:main-eq}, $K_\Delta \propto I$, does not follow from $K_\Delta |\psi\> \propto |\psi\>$ and the critical point condition in general, but does follow under certain assumptions that will be explained later in this section.

More generally, the result can be extended to the ground state on a finite circle and the thermal state on an infinite line.
In both cases, the entanglement entropy and the entanglement Hamiltonian are known \cite{calabrese2009entanglement,cardy2016entanglement}, so the proof strategy still works.
The difference is to replace the cross-ratio $\eta$ with the effective cross-ratio $\eta_{\text{eff}}$ where $\eta^L_{\text{eff}} = \frac{\sin(\pi x_{12}/L)\sin(\pi x_{34}/L)}{\sin(\pi x_{13}/L)\sin(\pi x_{24}/L)}$ for the ground state on a circle of length $L$
and $\eta^\beta_{\text{eff}} = \frac{\sinh(\pi x_{12}/\beta)\sinh(\pi x_{34}/\beta)}{\sinh(\pi x_{13}/\beta)\sinh(\pi x_{24}/\beta)}$ for the thermal state on an infinite line of inverse temperature $\beta$.

We now discuss the subtle relation between the operator equation $K_\Delta = \frac{c}{3} h(\eta)$ and the vector equation $K_\Delta |\psi\> = \frac{c}{3} h(\eta) |\psi\>$.
We first present an argument that shows their equivalence for quantum field theories, then discuss the difference in their approximate versions.
It is clear that the operator equation implies the vector equation and typically the other way does not hold for finite dimensional Hilbert spaces.
One extreme example for finite dimensional Hilbert spaces is that $K_X = - \log \rho_X$ could have singularities when $\rho_X$ has zero eigenvalues, while $K_X |\psi\>$ is still well-defined with continuity, because $\lim_{\lambda \to 0} \lambda \log \lambda = 0$.
On the other hand, for quantum field theories, due to the Reeh-Schlieder theorem \cite{reeh1961bemerkungen}, $|\psi\> \in \cH_{ABCD}$ is cyclic, which means $O_D|\psi\>$ is dense in $\cH_{ABC}$ for $O_D$ supported on $D$.
Because $K_\Delta$ is only supported on $\cH_{ABC}$,
  $K_\Delta$ and $O_D$ commutes,
  and we have $K_\Delta O_D |\psi\> = \frac{c}{3} h(\eta) O_D |\psi\>$.
Since $O_D|\psi\>$ is dense, this implies the operator equation $K_\Delta = \frac{c}{3} h(\eta)$.

In the approximate case, one might hope that $K_\Delta \approx \frac{c}{3} h(\eta)$ in the operator norm and $K_\Delta |\psi\> \approx \frac{c}{3} h(\eta) |\psi\>$ in the vector norm.
Numerically, we observe that the approximate operator equation does not hold, while the approximate vector equation holds.
This suggests that the vector equation is more valid because it is stable against perturbation.
Nevertheless, we suspect that the operator equation remains stable under a different norm which requires further investigation.


\section{Local-to-global implications}
In the last section, we established that $K_\Delta = \frac{c}{3} h(\eta)$ for any consecutive intervals $A, B, C$.
In this section, we apply this relation to intervals with integer-valued endpoints which can be associated with a lattice model.
The benefit of having a lattice model is that it allows for concrete numerical verifications.
We demonstrate that several quantities can be expressed as a sum of local operators.
First, we prove that the EH of an interval can be expressed as a sum of 1-site and 2-site EHs from Eq.~\eqref{eq:main-eq}.
This implies that the reduced density matrix on an arbitrary interval can be reconstructed from the 2-site reduced density matrix.

We first introduce the notation.
To simplify the presentation, we define the reduced EH as $\wtK_X \coloneqq K_X - \<\psi|K_X|\psi\>$, which is the EH shifted by a constant so that $\<\psi|\wtK_X|\psi\> = 0$.
This removes the constant term in Eq.~\eqref{eq:EH}, which implies that $\wtK_\Delta = 0$.
The reduced EH of an interval $[a, b]$ is denoted as $\wtK_{[a, b]}$.
In the following discussion, we only consider intervals with $a, b \in \ZZ$, so that they can be associated with a lattice model.
The corresponding lattice model regards each interval of length $1$, $[a, a+1]$, as a single site\footnote{\version{Of course, these `sites' have infinite-dimensional Hilbert spaces, and this is just a rewriting of the continuum CFT information.  In what follows, we will use the same relation in systems where the Hilbert space of each site is replaced by a finite-dimensional Hilbert space.}}.
Therefore, the interval $[a, b]$ corresponds to $b-a$ sites.
From now on, we refer to the reduced EH simply as EH.

We now show that any EH of an interval can be written as a sum of 1-site and 2-site EHs,
  using Eq.~\eqref{eq:main-eq} recursively.
We begin with the example of a 3-site EH.
Take $A = [0,1], B = [1,2], C = [2,3]$. The equivalent form of Eq.~\eqref{eq:main-eq} $\wtK_\Delta = 0$ implies
\begin{equation}
  \wtK_{[0,2]} + \wtK_{[1,3]} - \frac{1}{4} \wtK_{[0,1]} - \frac{1}{4} \wtK_{[2,3]} - \frac{3}{4} \wtK_{[1,2]} - \frac{3}{4} \wtK_{[0,3]} = 0.
\end{equation}
Therefore, 
\begin{equation}
  \wtK_{[0,3]} = - \frac{1}{3} \wtK_{[0,1]} + \frac{4}{3} \wtK_{[0,2]} - \wtK_{[1,2]} + \frac{4}{3} \wtK_{[1,3]} - \frac{1}{3} \wtK_{[2,3]}.
\end{equation}
Similarly, for a 4-site EH, take $A = [0,1], B = [1,2], C = [2,4]$, and we have
\begin{align}
  \wtK_{[0,4]} &= - \frac{1}{2} \wtK_{[0,1]} + \frac{3}{2} \wtK_{[0,2]} - \wtK_{[1,2]} + \frac{3}{2} \wtK_{[1,4]} - \frac{1}{2} \wtK_{[2,4]} \nonumber\\
  &= - \frac{1}{2} \wtK_{[0,1]} + \frac{3}{2} \wtK_{[0,2]} - \frac{3}{2} \wtK_{[1,2]} + 2 \wtK_{[1,3]} \nonumber\\
  &\qquad\qquad - \frac{3}{2} \wtK_{[2,3]} + \frac{3}{2} \wtK_{[2,4]} - \frac{1}{2} \wtK_{[3,4]} \label{eq:EH4}.
\end{align}
More generally, for an $n$-site EH, take $A = [0,1], B = [1,2], C = [2, n]$, and we have
\begin{align}
  \wtK_{[0,n]} &= - \frac{n-2}{n} \wtK_{[0,1]} + \frac{2(n-1)}{n} \wtK_{[0,2]} - \wtK_{[1,2]} \nonumber\\
  &\qquad\qquad + \frac{2(n-1)}{n} \wtK_{[1,n]} - \frac{n-2}{n} \wtK_{[2,n]}.
\end{align}
By recursively expanding $\wtK_{[1,n]}$ and $\wtK_{[2,n]}$, we obtain
\begin{equation} \label{eq:EHn}
  \wtK_{[0,n]} = \sum_{j=-\infty}^\infty f_2(j+1) \wtK_{[j, j+2]} + f_1(j+\frac{1}{2}) \wtK_{[j, j+1]}
\end{equation}
where $f_2(x) = \frac{2x(n-x)}{n} 1_{[0, n]}$, $f_1(x) = \frac{-2x(n-x)+\frac{3}{2}}{n} 1_{[0, n]}$, and $1_{[0, n]}$ is the indicator function.
Therefore, any EH can be written as a sum of 1-site and 2-site EHs, which means the reduced density matrix on arbitrary intervals can be reconstructed from the 2-site reduced density matrix.
This could be viewed as a solution to the quantum marginal problem for CFTs where the Petz recovery map does not apply.

\begin{remark} \label{remark:relation-of-relation}
  We note that the decomposition of EH into a sum of 1-site and 2-site EHs is similar to Eq.~\eqref{eq:EH} in field theory, where EH is a sum of local terms $T_{00}(x)$.
  We also remark that there are different ways to decompose $\wtK_{[0,n]}$, yet all of them lead to the same expression.
  This nontrivial property implies certain consistency relations between $\wtK_\Delta = 0$ across different choices of $A, B, C$.
\end{remark}

\emph{The state contains the universal data---}
The same logic that leads to Eq.~\eqref{eq:main-eq} implies that the CFT Hamiltonian (and hence all of the CFT data) can be extracted from the groundstate.
In fact, the same is true of all 6 global conformal generators $H, P, D, M_{10}, C_0, C_1$ which are the Hamiltonian, momentum, dilatation, boost, and special conformal generators.
We first find their corresponding expressions for CFT ground states when Eq.~\eqref{eq:EH} applies.
Because $H = \int_{-\infty}^{\infty} dx\, T_{00}(x)$, $M_{10} = \int_{-\infty}^{\infty} dx\, x T_{00}(x)$, $C_0 = \int_{-\infty}^{\infty} dx\, x^2 T_{00}(x)$,
  applying Eq.~\eqref{eq:EH}, we have
\begin{align}
  H &= \frac{1}{\pi} \sum_{j=-\infty}^{\infty} \left( \wtK_{[j, j+2]} - \wtK_{[j, j+1]} \right) \nonumber\\
    &= \frac{1}{\pi} \sum_{j=-\infty}^{\infty} \wtK'_{[j, j+2]},
  \label{eq:H-reconstruction} \\
  M_{10} &= \frac{1}{2 \pi}\sum_{j=-\infty}^{\infty} \left( (2j+2) \wtK_{[j, j+2]} - (2j+1) \wtK_{[j, j+1]}  \right) \nonumber\\
         &= \frac{1}{\pi}\sum_{j=-\infty}^{\infty} (j+1) \wtK'_{[j, j+2]},
  \label{eq:M10-reconstruction} \\
  C_0 &= \frac{1}{\pi} \sum_{j=-\infty}^{\infty} \left( (j+1)^2 \wtK_{[j, j+2]} - (j^2+j+1) \wtK_{[j, j+1]}  \right) \nonumber\\
      &= \frac{1}{\pi} \sum_{j=-\infty}^{\infty} \left( (j+1)^2 \wtK'_{[j, j+2]} - \frac{1}{2} \wtK_{[j, j+1]} \right),
  \label{eq:K0-reconstruction}
\end{align}
where $\wtK'_{[j, j+2]} = \wtK_{[j, j+2]} - \frac{1}{2} \wtK_{[j, j+1]} - \frac{1}{2} \wtK_{[j+1, j+2]}$ is introduced to simplify the equations.
Because $P = \ii [M_{10}, H]$, $D = \frac{\ii}{2} [C_0, H]$, $C_1 = \ii [C_0, M_{10}]$, we have
\begin{align}
  P &= \frac{\ii}{\pi^2} \sum_{j=-\infty}^{\infty} \left[\wtK'_{[j+1, j+3]}, \wtK'_{[j, j+2]}\right],
  \label{eq:P-reconstruction} \\
  D &= \frac{\ii}{\pi^2}  \sum_{j=-\infty}^{\infty} \bigg((j+\frac{3}{2}) \left[\wtK'_{[j+1, j+3]}, \wtK'_{[j, j+2]}\right] \nonumber\\
  &\qquad\qquad - \frac{1}{4} \left[\wtK_{[j, j+1]}, \wtK'_{[j, j+2]}\right] \nonumber\\
  &\qquad\qquad - \frac{1}{4} \left[\wtK_{[j+1, j+2]}, \wtK'_{[j, j+2]}\right] \bigg),
  \label{eq:D-reconstruction} \\
  C_1 &= \frac{\ii}{\pi^2}  \sum_{j=-\infty}^{\infty} \bigg( (j+1)(j+2) \left[\wtK'_{[j+1, j+3]}, \wtK'_{[j, j+2]}\right] \nonumber\\
  &\qquad\qquad - \frac{j+1}{2} \left[\wtK_{[j, j+1]}, \wtK'_{[j, j+2]}\right] \nonumber\\
  &\qquad\qquad - \frac{j+1}{2} \left[\wtK_{[j+1, j+2]}, \wtK'_{[j, j+2]}\right] \bigg).
  \label{eq:K1-reconstruction}
\end{align}
This constructs the 6 global conformal generators from 1-site and 2-site EHs.
\version{We use the term reconstructed Hamiltonian to refer to the output of the process of taking a state and constructing an associated Hamiltonian as in \eqref{eq:H-reconstruction}. This state could be on a lattice or in the continuum.}

We make two comments on the expression for $P$.
First, a similar expression was considered in \cite[Sec II.C]{milsted2017extraction}.
The idea is to define the momentum density $p_j = \ii [h_j, h_{j-1}]$ in terms of the Hamiltonian density $h_j$ where $H = \sum h_j$.
One challenge in implementing this idea in practice is that the Hamiltonian $H$ is only equal to the CFT Hamiltonian $H_{CFT}$ up to a multiplicative factor.
Consequently, the momentum density $p_j$ is determined only up to a multiplicative factor.
In our work, we fix this multiplicative factor and provide a better theoretical understanding.
Second, there is a no-go theorem that prevents expressing the momentum operator as a sum of local operators \cite[Corollary 6.1]{ranard2022converse} which seemingly contradicts our expression above.
There are several ways to reconcile this apparent contradiction.
One is to note that our expression holds exactly only for lattice models with an infinite local Hilbert space dimension, whereas the no-go theorem applies to models with a finite local Hilbert space dimension.
Another way to reconcile this is to recognize that, for models with a finite local Hilbert space dimension, $e^{-\ii P}$ is not precisely equal to the lattice translation operator \cite{cheng2022lieb}.
Instead, they are only approximately equal at low energies, as we verify numerically below.

We further comment that there are additional commutation relations between the generators which are not utilized in this work,
  such as $[M_{10}, P] = - \ii H$ and $[H, P] = 0$.
In the context of field theories, this phenomenon where higher order commutator of EHs are related to linear combinations of EHs
  can be understood from the operator product expansion (OPE) of $T_{zz}(t, x)$.
First, because EH is a sum of $T_{zz}(t=0, x)$ and $T_{\bar z \bar z}(t=0, x)$,
  the commutator of EHs is a sum of the commutators of $T_{zz}(t=0, x)$ and $T_{\bar z \bar z}(t=0, x)$.
Second, the commutator of $T_{zz}(t=0, x)$ is generated by the singular part of the OPE of $T_{zz}$.
Finally, the singular part of the OPE of $T_{zz}$ is generated by $T_{zz}$ and its descendent $\partial_z T_{zz}$.
The second and the final part can be summarized into the following expression \cite[Equation (B23)]{zou2022modular}
\begin{align}
  [T_{zz}(0,x_1), T_{zz}(0,x_2)] &= \frac{\ii \pi c}{6} \partial_{x_1}^3 \delta(x_1 - x_2) \nonumber\\
  &\hspace*{-1em} + 4 \pi \ii T_{zz}(0, x_2) \partial_{x_1} \delta(x_1 - x_2) \nonumber\\
  &\hspace*{-1em} - 2 \pi \ii \partial_{z} T_{zz}(0, x_2) \delta(x_1 - x_2).
\end{align}
We leave these further relations between the higher order commutators of EHs and the linear combinations of EHs for future study.

Having defined the 6 conformal generators using 1-site and 2-site EHs, we now show that they have the expected scaling from real-space RG.
We first study the case of the Hamiltonian $H$.
To perform real-space RG, we consider a new system which blocks 2 sites in the original system into 1 site.
We then compare the two reconstructed Hamiltonians using Eq.~\eqref{eq:H-reconstruction}, which are
\begin{align}
  H_1 &= \frac{1}{\pi} \sum_{j=-\infty}^\infty \wtK_{[j, j+2]} - \wtK_{[j, j+1]} \\
  H_2 &= \frac{1}{\pi} \sum_{j=-\infty}^\infty \wtK_{[2j, 2j+4]} - \wtK_{[2j, 2j+2]}.
\end{align}
Note that the state we used for the reconstructions is the same.
  The only difference is the size of the block.
By expanding the 4-site EH as a sum of 1-site and 2-site EH using Eq.~\eqref{eq:EH4},
  we have $H_2 = 2 H_1$.
Similarly, one can define $P_2$ and show that $P_2 = 2 P_1$, $D_2 = D_1$, $(M_{10})_2 = (M_{10})_1$, $(C_0)_2 = \frac{1}{2} (C_0)_1$, and $(C_1)_2 = \frac{1}{2} (C_1)_1$.
These are precisely the scalings of the generators for CFT under the transformation $x \to x/2$, $t \to t/2$.

It is perhaps not surprising that the scaling property holds because the construction is motivated by the field theories.
However, what is surprising is that the derivation of the scaling property only uses the condition \eqref{eq:main-eq}, which is independent from the field theory description of CFTs.
This observation supports the conjecture that states satisfying Eq.~\eqref{eq:main-eq} are the CFT ground states.



\section{Numerical tests}
We now move on to numerics and verify that Eqs.~\eqref{eq:main-eq}, \eqref{eq:main-eq-with-ratio}, \eqref{eq:H-reconstruction} and \eqref{eq:P-reconstruction} hold approximately for critical lattice models.
First, we test the validity of the main Eqs.~\eqref{eq:main-eq} and \eqref{eq:main-eq-with-ratio} on small sizes of four different critical lattice models and on large sizes of free fermions.
We find that these equations hold approximately for all the models we consider and the error decreases as the size increases.
Then, we test the reconstruction of the Hamiltonian and momentum operator Eqs.~\eqref{eq:H-reconstruction} and \eqref{eq:P-reconstruction} on small sizes of the four models.

The critical lattice models we consider are the critical transverse field Ising model, critical three-state Potts model, XX model, and Heisenberg model\footnote{\version{These models are all quantum critical and their universal properties are described respectively by the following conformal field theories: Ising model with $c=1/2$; Potts model with $c=4/5$; compact scalar CFT with radius $1$ and $1/\sqrt{2}$}.} defined as:
\begin{align}
  H_{\text{Ising}} &= \sum_i -Z_i -X_i X_{i+1} \\
  H_{\text{Potts}} &= \sum_i -\cZ_i -\cZ^\dagger_i -\cX_i \cX^\dagger_{i+1} -\cX^\dagger_i \cX_{i+1} \\
  H_{\text{XX}} &= \sum_i X_i X_{i+1} + Y_i Y_{i+1} \\
  H_{\text{Heisenberg}} &= \sum_i X_i X_{i+1} + Y_i Y_{i+1} + Z_i Z_{i+1}
\end{align}
where $X, Y, Z$ are the Pauli matrices and $\cX, \cZ$ are the qutrit Pauli matrices
\begin{equation}
  \cX = \begin{pmatrix}
    0 & 1 & 0 \\
    0 & 0 & 1 \\
    1 & 0 & 0
  \end{pmatrix}, \quad
  \cZ = \begin{pmatrix}
    1 & 0 & 0 \\
    0 & \omega & 0 \\
    0 & 0 & \omega^2
  \end{pmatrix}, \quad
  \omega = e^{2 \pi \ii / 3}
\end{equation}

\Cref{table:test-main-eq} lists the errors associated with Eqs.~\eqref{eq:main-eq} and \eqref{eq:main-eq-with-ratio}.
The error in Eq.~\eqref{eq:main-eq} is defined as the norm of the component in $K_\Delta |\psi\>$ orthogonal to $|\psi\>$, $\Big\lvert K_\Delta |\psi\> - \<\psi|K_\Delta|\psi\> |\psi\>\Big\rvert$,
which is also the standard deviation of $K_\Delta$,
$\sqrt{\<K_\Delta^2\> - \<K_\Delta\>^2}$.
The error in Eq.~\eqref{eq:main-eq-with-ratio} is defined as the norm $\Big\lvert K_\Delta |\psi\> - \frac{c}{3} h(\eta) |\psi\>\Big\rvert$.
We computed these errors for ground states on circles with circumferences $L = 4, 8, 12$, where $(A, B, C)$ have lengths $(1,1,1), (2,2,2),$ and $(3,3,3)$, respectively. In all cases, $\eta = 1/2$.
We observe that except for the accidental case where the XX model and Heisenberg model with circumference $4$ have $0$ errors, the error decreases as the system size increases.
This result is consistent with the intuition that the lattice model approximates the CFT better in the IR as the system size increases.

\begin{table}
  \centering
  \footnotesize
  \begin{tabular}{c @{\kern20pt} ccc @{\kern20pt} ccc}
  \toprule
  \multirow2*{$L$}
   & \multicolumn{3}{c}{\kern-20pt       Error in Eq.~\eqref{eq:main-eq}}
   & \multicolumn{3}{c}{\kern-\tabcolsep Error in Eq.~\eqref{eq:main-eq-with-ratio}} \\
              & 4      & 8      & 12     & 4      & 8      & 12     \\
  \midrule
  Ising Model & 0.0282 & 0.0090 & 0.0057 & 0.0288 & 0.0090 & 0.0057 \\
  Potts Model & 0.0422 &        &        & 0.0434 &        &        \\
  XX Model    & 0      & 0.0399 & 0.0120 & 0.0486 & 0.0416 & 0.0125 \\
  Heisenberg  & 0      & 0.0562 & 0.0279 & 0.0566 & 0.0583 & 0.0286 \\
  \bottomrule
  \end{tabular}
  \caption{The error in Eq.~\eqref{eq:main-eq} $\Big\lvert K_\Delta |\psi\> - \<\psi|K_\Delta|\psi\> |\psi\>\Big\rvert$ and the error in Eq.~\eqref{eq:main-eq-with-ratio} $\Big\lvert K_\Delta |\psi\> - \frac{c}{3} h(\eta) |\psi\>\Big\rvert$ for ground states on circles with circumferences $L = 4, 8, 12$.}
  \label{table:test-main-eq}
\end{table}

\begin{figure}
  \centering
  \includegraphics[width=0.49\linewidth]{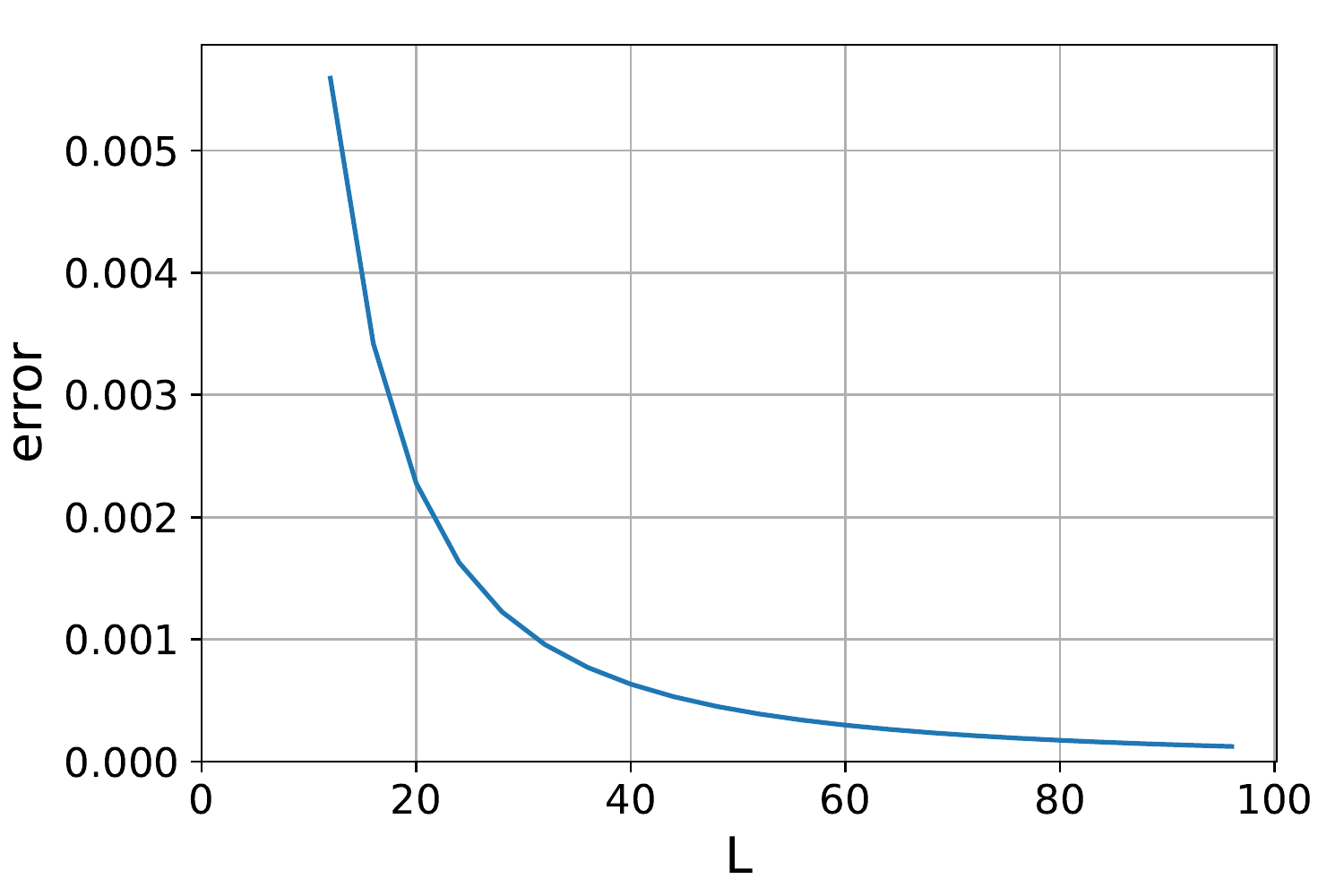}
  \includegraphics[width=0.49\linewidth]{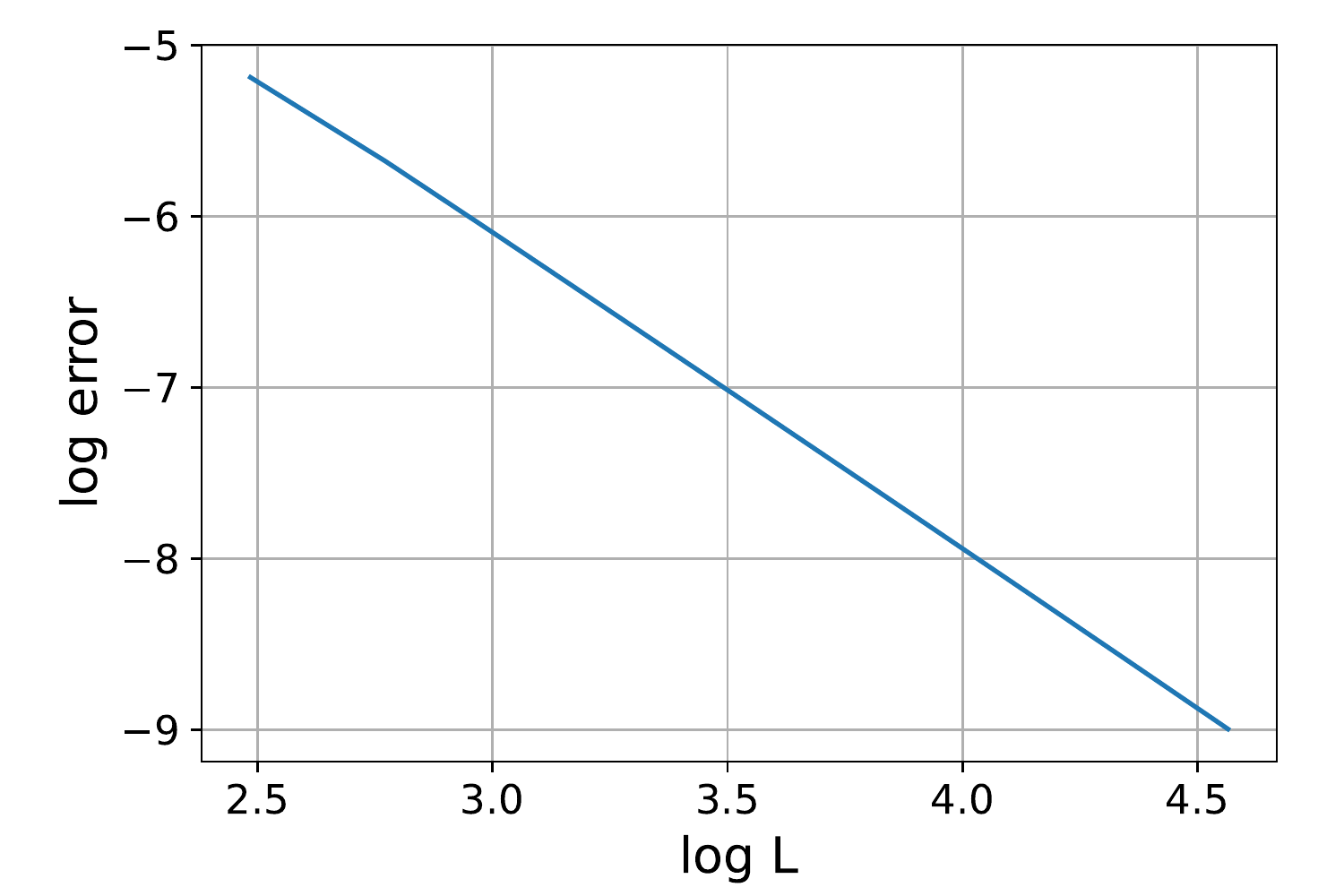}
  \caption{The error in Eq.~\eqref{eq:main-eq} $\Big\lvert K_\Delta |\psi\> - \<\psi|K_\Delta|\psi\> |\psi\>\Big\rvert$ for the ground state of the critical free fermions. The circle is partitioned into four equal sized intervals, where the lengths of $A, B, C$ are $L/4$ and has cross ratio $\eta = 1/2$. The error roughly scales as $1/L^2$.}
  \label{fig:test-main-eq2}
\end{figure}

\Cref{fig:test-main-eq2} shows the error in Eq.~\eqref{eq:main-eq} for free fermions across various system sizes.
We again observe that the error decreases as the system size increases which roughly scales as $1/L^2$.
Simulating free fermions on a large system size is feasible due to their lower complexity \cite{peschel2009reduced,fagotti2010entanglement}.

\begin{figure}
  \centering
  \includegraphics[width=0.49\linewidth]{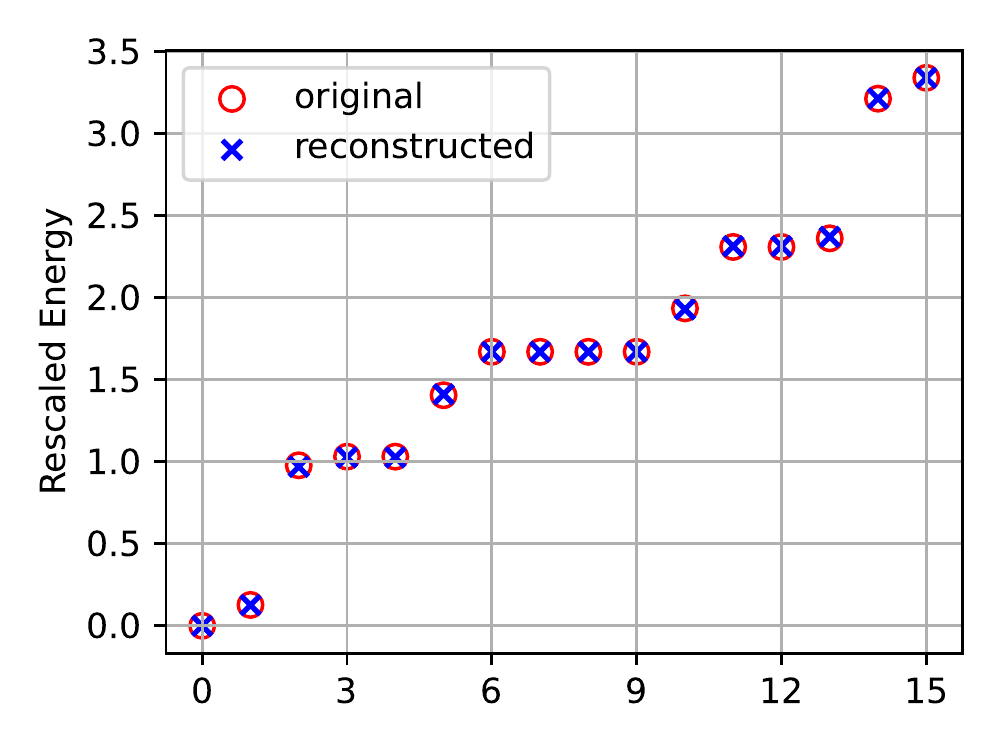}
  \includegraphics[width=0.49\linewidth]{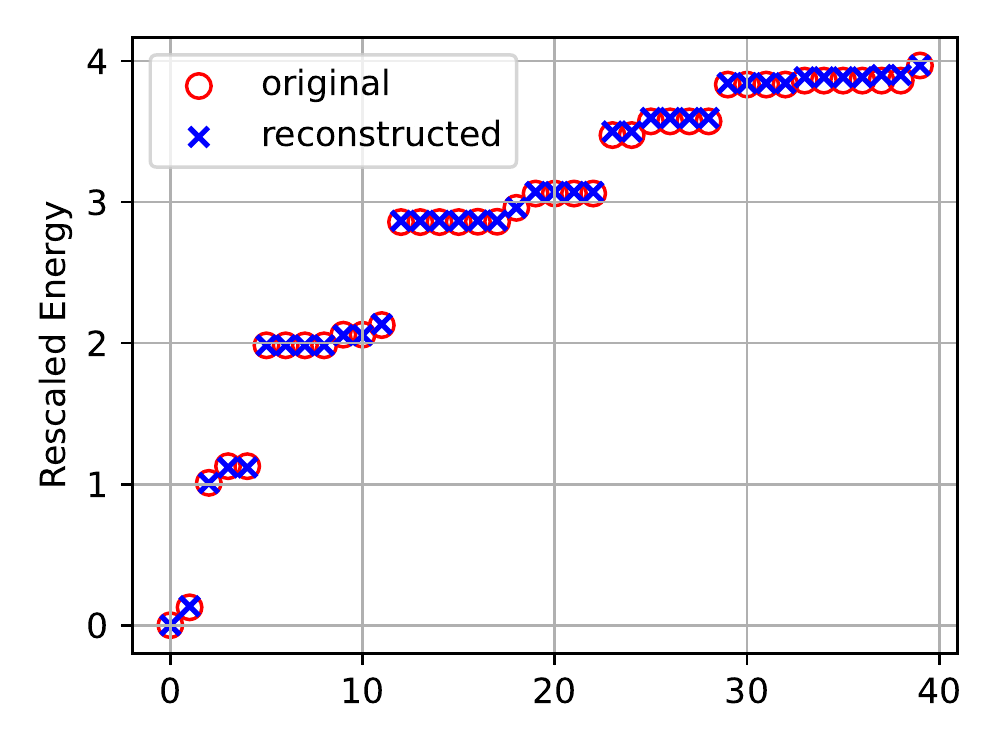}\\
  \includegraphics[width=0.49\linewidth]{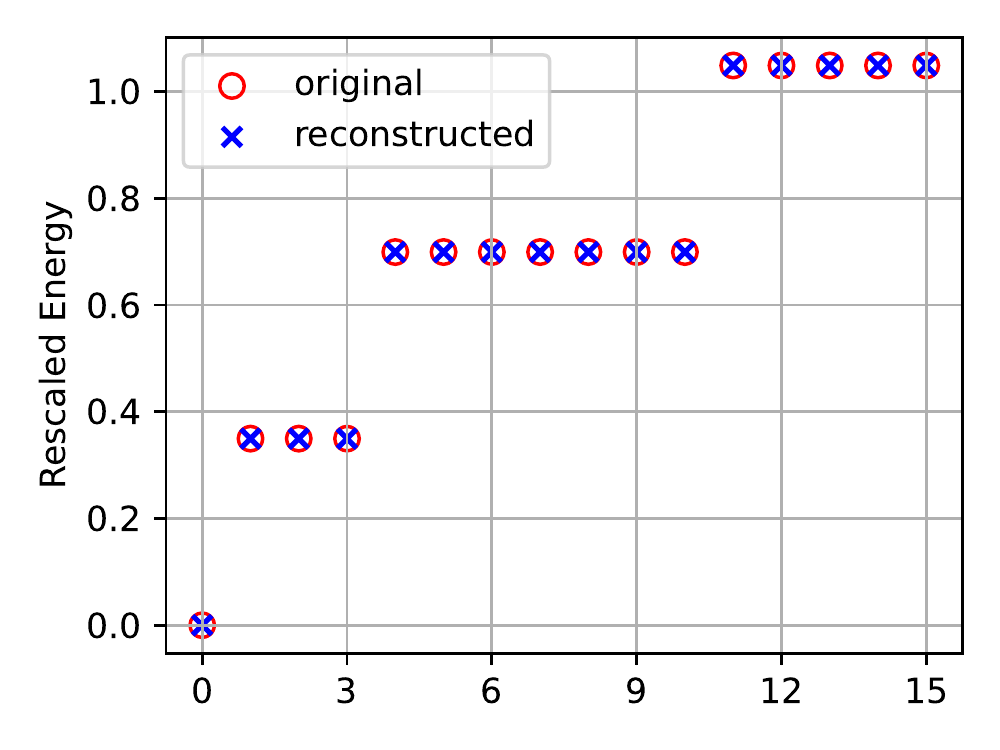}
  \includegraphics[width=0.49\linewidth]{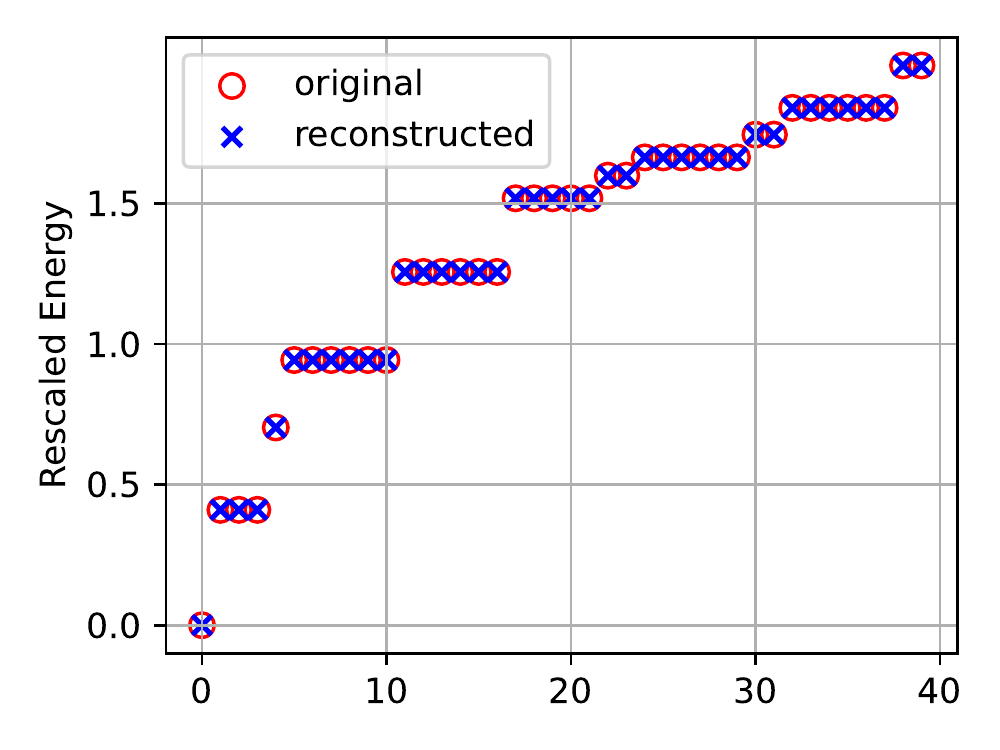}
  \caption{Spectra of the original Hamiltonian and the reconstructed Hamiltonian,
    where the states are ordered by energy along the X-axis.
    Top left: Ising model with $L = 4$.
    Top right: Lowest $40$ eigenvalues for Ising model with $L = 12$.
    Bottom left: Heisenberg model with $L = 4$.
    Bottom right: Lowest $40$ eigenvalues for Heisenberg model with $L = 12$.
  }
  \label{fig:test-reconstructed-H}
\end{figure}

\begin{figure}
  \centering
  \includegraphics[width=0.49\linewidth]{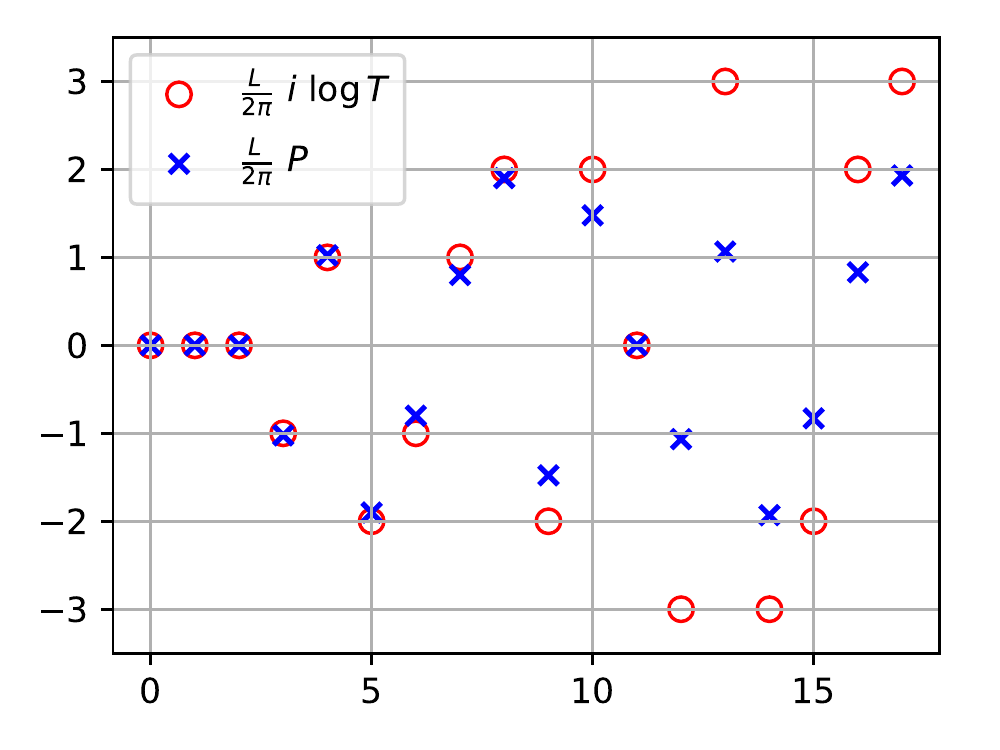}
  \includegraphics[width=0.49\linewidth]{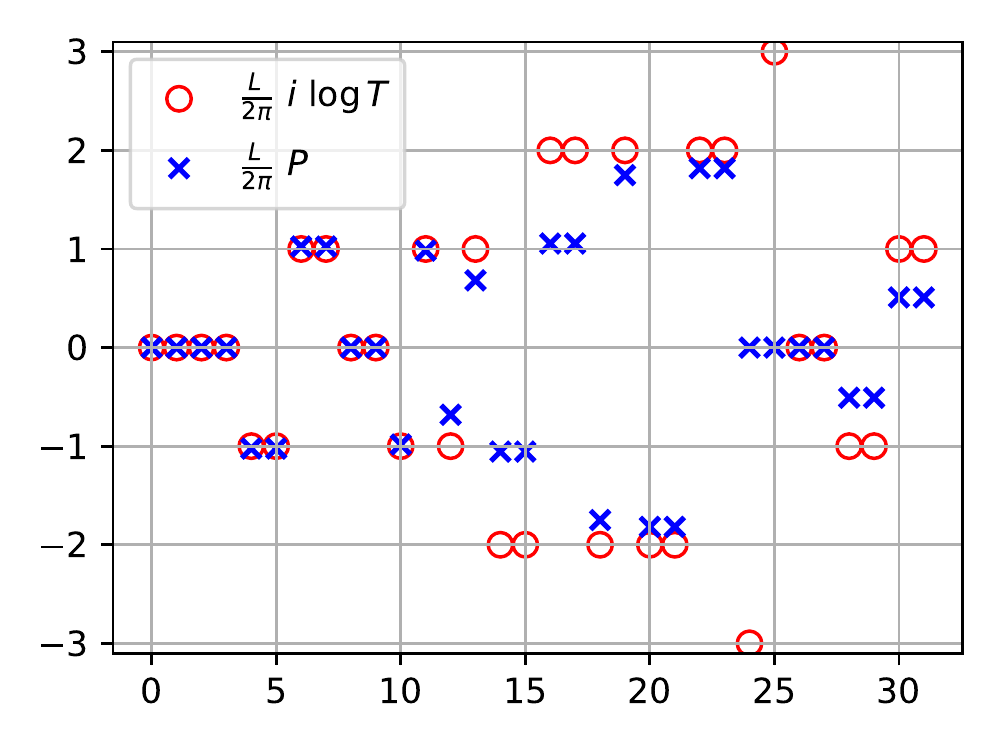} \\
  \includegraphics[width=0.49\linewidth]{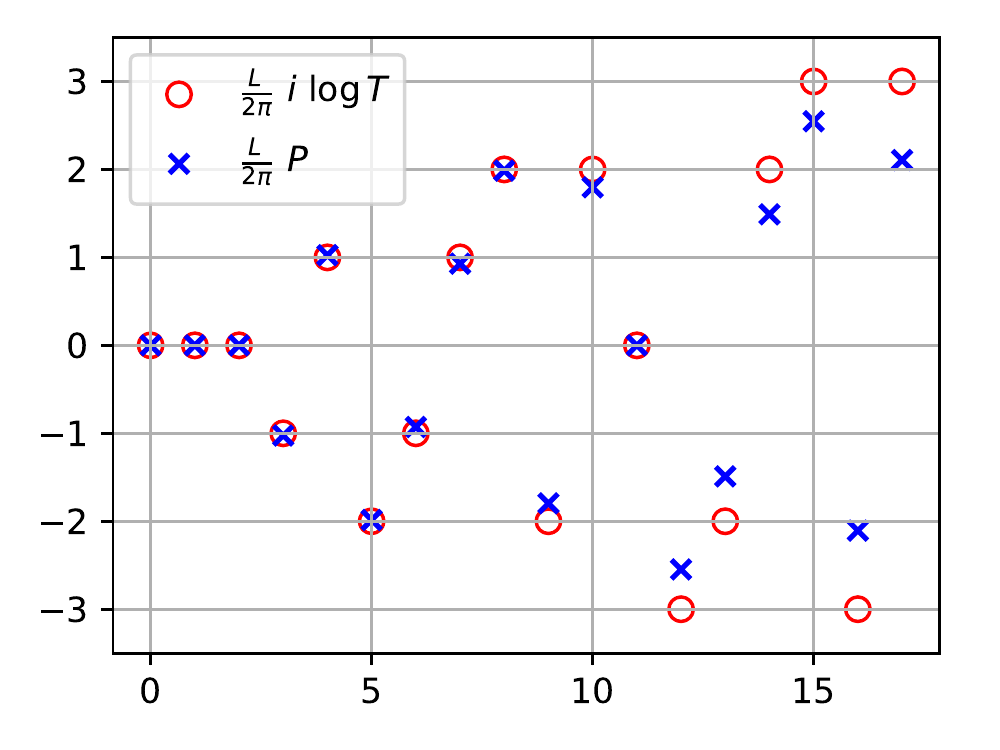}
  \includegraphics[width=0.49\linewidth]{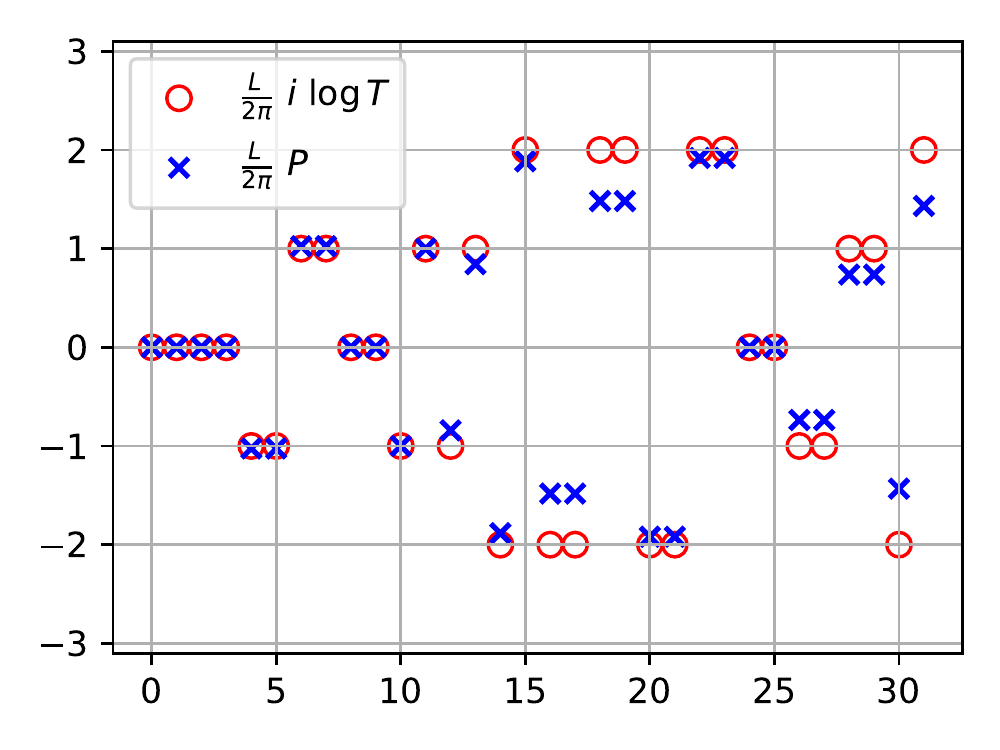}
  \caption{Spectra of $\ii \log T$ and the reconstructed momentum operator $P$,
    where $T$ is the translation operator by 1 site.
    The states are ordered by energy along the X-axis.
    In the infinite-size limit, this order corresponds to the order of the scaling dimension of the corresponding operator from the state-operator correspondence.
    For the small system sizes shown here, these orders are not yet the same.
    This is why the order changes in the figure when the system size changes.
    Top left: Ising model with $L = 8$.
    Bottom left: Ising model with $L = 12$.
    Top right: Potts model with $L = 6$.
    Bottom right: Potts model with $L = 8$.
  }
  \label{fig:test-reconstructed-P}
\end{figure}

We now test the reconstructed Hamiltonian and momentum on a circle in Eqs.~\eqref{fig:test-reconstructed-H} and \eqref{fig:test-reconstructed-P}
where
\begin{align}
  H_{rec} &= \frac{\sin (2 \pi / L)}{2L \sin^2 (\pi / L)} \sum_{j=0}^{L-1} \wtK_{[j, j+2]} - \wtK_{[j, j+1]} \label{eq:H-finite-reconstruction}\\
  P_{rec} &= \frac{\ii}{\pi^2} \sum_{j=0}^{L-1} \left[\wtK'_{[j+1, j+3]}, \wtK'_{[j, j+2]}\right]. \label{eq:P-finite-reconstruction}
\end{align}
\footnote{\version{The analogous formula for the exact reconstructed momentum is \begin{align*} P_{rec} &= \frac{\sin^3 (2 \pi / L) \pi}{8L \sin^4 (\pi / L)} \frac{i}{\pi^2} \sum_{j=0}^{L-1} [\wtK_{[j+1, j+3]}, \wtK_{[j, j+2]}] \\ &\hspace{6em} - \left(1-\frac{2 \sin^2(\pi / L)}{\sin^2(2 \pi / L)}\right) [\wtK_{[j, j+2]}, \wtK_{[j, j+1]}] \\ &\hspace{6em} - \left(1-\frac{2 \sin^2(\pi / L)}{\sin^2(2 \pi / L)}\right) [\wtK_{[j+1, j+2]}, \wtK_{[j, j+2]}] \end{align*}}}
The factor $\frac{\sin (2 \pi / L)}{2L \sin^2 (\pi / L)}$ in $H_{rec}$ comes from the coefficient obtained when expressing EH as an integral of $T_{00}(x)$.
On the other hand, the factor $\frac{\ii}{\pi^2}$ in $P_{rec}$ is based solely on the understanding at the infinite-size limit in Eq.~\eqref{eq:P-reconstruction}.
Despite this, as we will see, the reconstructed Hamiltonian and momentum agree excellently. \footnote{\version{Note that $H_{rec}$ is equal to $H_{CFT}$ since both have the same expression in terms of $T$. On the contrary, because we put the system on a circle $P_{rec}$ is only equal to $P_{CFT}$ in the thermodynamic limit.}}

Before presenting the result, we first remark that the field theory Hamiltonian and momentum satisfy
\begin{align}
  H_{CFT}\, |\Delta, s\> &= \frac{2\pi}{L} \left(\Delta - \frac{c}{12}\right) |\Delta, s\> \\
  P_{CFT}\, |\Delta, s\> &= \frac{2\pi}{L} s\, |\Delta, s\>,
\end{align}
where $|\Delta, s\>$ is the
image under the state-operator correspondence
of a scaling operator
of dimension $\Delta$ and spin $s$.
We emphasize that $H_{rec}$ is equal to $H_{CFT}$ up to a constant shift so that $H_{rec}$ has ground state energy $0$, i.e. $H_{rec} = H_{CFT} - E_0$
\footnote{The constant $-\frac{2\pi}{L}\frac{c}{12}$ can be recovered by using the EHs of the ground state on an infinite system instead of the EHs of the length $L$ finite system, which we have tested for critical free fermions. However, we do not discuss this aspect further because it is generally not possible to know the EHs on an infinite system directly.}.
In particular, the multiplicative factor is fixed, meaning that the scaling dimensions $\Delta$ can be obtained without the rescaling required if one directly uses the spectrum of an arbitrary critical Hamiltonian.

\Cref{fig:test-reconstructed-H} compares the spectra of the original Hamiltonian to the spectrum of the reconstructed Hamiltonian at low energy.
The reconstructed Hamiltonian $H_{rec}$ is rescaled by $\frac{L}{2\pi}$ and the original Hamiltonian is rescaled to fit $H_{rec}$.
We observe the spectra have excellent agreement even at the small system size where $L=4$.\footnote{\version{We emphasize that this is a pleasant surprise: although we use field theory to derive our formula for the reconstructed Hamiltonian, its form is numerically effective even on quite small lattices.
It would be interesting to understand why it works so well.

Locality clearly plays an important role.
We note that the ideal Hamiltonian reconstructed from the saddle point of the entropy function $S_\Delta$
and the model Hamiltonians we study
(Eqs.~24-27) both involve at most 2-site operators.  This means that the artificial model Hamiltonian can already be close to the ideal reconstructed CFT Hamiltonian.}}

\Cref{fig:test-reconstructed-P} compares the spectrum of $\ii \log T$ to the spectrum of the reconstructed momentum at low energy, where $T$ is the translation operator by 1 site.
We test if $T \approx e^{-\ii P_{rec}}$.
Again both spectra are rescaled by $\frac{L}{2\pi}$ and we expect both to take integer values.
We observe the spectra agree at low energy and the agreement improves as the system size increases.

We provide three technical comments on this test.
First, we note that in lattice models, it is not generally true that $T = e^{-\ii P_{CFT}}$, for example, in the Heisenberg and XX models.
This is because $\ii \log T$ may have a constant shift $s_0$, where $\ii \log T |\psi\> = (s_0 + \frac{2\pi}{L} s) |\psi\>$.
As an example, if we instead define the transverse field Ising model as $H_\text{Ising} = \sum_i - Z_i + X_i X_{i+1}$,
  then the new ground state is obtained by applying $Z$ on the even sites of the usual ground state.
This results with a shift $s_0 = \pi$.
Sometimes this shift cannot be removed easily due to anomaly \cite{cheng2022lieb}.
This is why we did not show the reconstruction in the case of Heisenberg and XX models for simplicity.
Next, we remark on two surprising properties of $P_{rec}$.
One is that $\frac{L}{2\pi} P_{rec}$ is close to taking integer values at low energy.
The other is that it is often the case that $-\pi \le P_{rec} \le \pi$.
Both properties are expected when knowing $T \approx e^{-\ii P_{rec}}$.
However, from Eq.~\eqref{eq:P-finite-reconstruction} alone, it is unclear why this happens; we leave this for future research.
Finally, although we lack a full analytic understanding of the factor $\frac{\ii}{\pi^2}$ in $P_{rec}$, it works well numerically.
The precise nature of this factor on a finite system, and whether this is a correct choice or merely a coincidence, is left for further exploration.

\emph{Further directions---}
The further directions are sorted based on some subjective measure of attractiveness.

(a) We show that CFT ground states satisfy the condition in Eq.~\eqref{eq:main-eq}.
Motivated by entanglement bootstrap, we ask whether the converse is also true, i.e. if a state satisfying the condition can be interpreted as a CFT ground state.
Additionally, we would like the statement to be robust, useful even when the state only satisfies the condition approximately.

(b) We proposed various formulas that hold exactly when the state is a CFT ground state.
However, for the ground state of critical lattice models, the formula only holds approximately.
What is the convergence behavior and the finite size scaling?

(c) Relatedly, one may wonder if we allow combinations of not only 1-site and 2-site EH but also 3-site EH,
which combination has the best approximation to the Hamiltonian and the momentum.
Furthermore, what if we include EH on larger sites?
Could this lead to a sequence of reconstructions that converges to the actual operator?
Finding a good approximation is crucial for obtaining a better estimation of the OPE based on the method discussed in \cite{zou2020conformal}.

(d) Can we reproduce the key results in CFT directly using quantum information? For example, can we show that all nontrivial states must have $c \ge \frac{1}{2}$? Can we show the ground state energy is $-\frac{2 \pi}{L} \frac{c}{12}$ on a circle with circumference $L$? Can we show that the reconstructed momentum operator should be approximately integer valued at low energy and has a norm roughly below $\pi$?

(e) The appearance of $h(\eta)$ in Eq.~\eqref{eq:main-eq-with-ratio} suggests that $\eta$ can be interpreted as a probability. Could we find a physical meaning for this observation?

(f) We provide an information theoretic criterion for CFT ground state.
Can we transform this criterion into an algorithm that searches for CFTs by screening states that satisfy Eq.~\eqref{eq:main-eq}?
We believe the answer is yes and plan to explore this aspect in an upcoming paper.

(g) We provide an entropy criterion that appears to describe ground states for 1+1D unitary CFTs.
We suspect a similar formula exists for 1+1D chiral CFTs, d+1D CFTs, \version{and perhaps complex CFTs \cite{Gorbenko:2018ncu,Gorbenko:2018dtm}.}

(h) We show that CFT ground states are critical points of the function $S_\Delta$, with a value proportional to the central charge. What is its relation to the Zamolodchikov $c$-function \cite{zamolodchikov1986irreversibility} and the entropic $c$-function \cite{casini2007c}? For example, it is known that the second and third-order expansion of the Zamolodchikov $c$-function near the CFT point contains the CFT data, including the scaling dimensions and the OPEs. Does the same hold for the entropy function we proposed?

(i) We showed that 1+1D phases at RG fixed points with Lorentz symmetry satisfy Eq.~\eqref{eq:main-eq}. What happens to other RG fixed points without Lorentz symmetry, such as those with dynamical critical exponent $z \ne 1$. In the case of free fermion models with $z \in \ZZ^{> 0}$, the formula continues to hold. When $z$ is even, the ground state is simply a product state, and when $z$ is odd, the ground state is same as the ground state for $z = 1$, which is a CFT.



\section{Acknowledgments}
We thank Isaac Kim and Hao-Chung Cheng for suggesting a simple proof to compute the first-order derivative of entanglement entropy in Eq.~\eqref{eq:entropy-first-order}.
We thank Meng Cheng and Yijian Zhou for clarifying why in general $T \ne e^{-\ii P}$.
We thank Bowen Shi and Daniel Ranard for helpful discussions.
This work was supported in part by funds provided by the U.S. Department of Energy (D.O.E.) under the cooperative research agreement DE-SC0009919, and by the Simons Collaboration on Ultra-Quantum Matter, which is a grant from the Simons Foundation (652264, JM).
TCL would like to thank the assistance of ChatGPT in improving parts of the paper for clarity, including this sentence.

\bibliographystyle{ucsd.bst}
\bibliography{references.bib}
\clearpage

\appendix
\section{Supplemental Material}

\subsection{Proof of Eq.~\eqref{eq:entropy-first-order}} \label{sec:proof-of-entropy-first-order}

Let $|\psi\> \in \cH_A \otimes \cH_B$ be a pure state.
We now show the first order derivative of entanglement entropy $S_A(|\psi\>)$ claimed in Eq.~\eqref{eq:entropy-first-order}.
Because $d \rho_A = \Tr_B(|\psi\>\<d \psi|) + \Tr_B(|d \psi\>\<\psi|)$,
it is sufficient to show
\begin{equation}
  d S(\rho_A) = \Tr_A((-\log \rho_A) d \rho_A).
\end{equation}
Using product rule in calculus,
\begin{equation}
  d S(\rho) = d \Tr((-\log \rho) \rho) = \Tr((-\log \rho) d \rho) - \Tr(\rho\, d (\log \rho)).
\end{equation}
We suffice to show the second term on the RHS is $0$.
This is done by showing
\begin{equation} \label{eq:key-intermediate-step}
  \Tr(\rho\, d (\log \rho)) = \Tr(d \rho)
\end{equation}
where $\Tr(d \rho) = d \Tr(\rho) = 0$.

Because $\rho$ is hermitian, we can write $\rho = e^M$.
Equivalently, we need to show
\begin{equation}
  \Tr(e^M d M) = \Tr(d e^M).
\end{equation}
This is obtained by an equation from the perturbation theory of path integrals
\begin{equation}
  d e^M = \int_0^1 dt\, e^{tM} d M e^{(1-t)M}.
\end{equation}
We sketch the proof of the final equation:
\begin{align}
  d e^M
  &= \lim_{n \to \infty} d \prod_{i=1}^n e^{M/n} \\
  &= \lim_{n \to \infty} \sum_{i=1}^n e^{\frac{i-1}{n}M} \,(d e^{M/n})\, e^{\frac{n-i}{n}M} \\
  &= \lim_{n \to \infty} \sum_{i=1}^n e^{\frac{i-1}{n}M} \,(d M/n)\, e^{\frac{n-i}{n}M} \\
  &= \int_0^1 dt e^{tM} d M e^{(1-t)M}
\end{align}
where the third equality uses $e^{M/n} = I + M/n + O(1/n^2)$.
The higher order term $O(1/n^2)$ converges to $0$ as $n \to \infty$, because there are only $n$ terms.

More directly, the key step Eq.~\eqref{eq:key-intermediate-step} can also be obtained through the following equality \cite[Lemma 3.4]{sutter2017multivariate}
\begin{equation}
  d \log \rho = \int_{-\infty}^\infty dt\, \beta_0(t)\, \rho^{-\frac{1+\ii t}{2}} \,d \rho\, \rho^{-\frac{1-\ii t}{2}}
\end{equation}
where $\beta_0(t) = \frac{\pi}{2} (\cosh(\pi t) + 1)^{-1}$.
When substituting $d \log \rho$ in Eq.~\eqref{eq:key-intermediate-step}, we obtain
\begin{equation}
  \Tr(\rho\, d (\log \rho)) = \int_{-\infty}^\infty dt\, \beta_0(t) \Tr(d \rho) = \Tr(d \rho)
\end{equation}
using $\int_{-\infty}^\infty dt \beta_0(t) = 1$.

\subsection{No other relations}

\version{
Here we argue that relations of the form (1) or (3)
exhaust linear relations among single-interval entanglement Hamiltonians of CFT groundstates.
What we mean is that any other linear entropy formula derivable from (5) and (6)
is implied by our fixed-point equation (1) or (3).

The idea is the following.
The observation is that each $K_A$ corresponds to a piecewise quadratic function $\beta_A$.
We wish to identify all linear relations between these piecewise quadratic functions.
For convenience, divide up the line or circle into a set of atomic intervals (we'll revisit this below).
Our relation shows that $\beta_A$ for any region can be decomposed into a linear combination of one-site and two-site functions, meaning the functions associated with a single atomic interval or two atomic intervals.
So it is sufficient to show that
there is no linear dependence among one-site and two-site functions.  This is easy to check: $ a (x_3-x)(x-x_1) + b( x_3 -x )(x-x_2) + c ( x_4 - x) (x-x_2) = 0 $
for all $x \in [x_1,x_2]$ implies $ a=b=c=0$.

If instead we wish to work in the continuum, suppose there
is some relation that is not of our form.
Use the intervals appearing in the relation as the atomic intervals and repeat the above argument.


We remark in passing that this question of other relations can be described as a cohomology problem in functional analysis.
Consider a formal vector space $V$ spanned by states labelled by intervals on the line or circle $\ket{A}$, with no relations.  There is map $\phi$ from $V$ to the Hilbert space of piecewise quadratic functions $\phi: \ket{A} \to \beta_A(x)$.  The kernel of the map $\phi$ is spanned by all relations that follow from (5) and (6).  There is map $\psi$ into $V$ determined by relations of the form (1) or (3).  The previous paragraph shows that there is no cohomology at $V$ -- the kernel of $\phi$ is the image of $\psi$.
The fact noted above \Cref{remark:relation-of-relation} about the associativity of decomposition of $\wtK_{[0,n]}$ implies that there are relations among our relations, meaning that this chain complex can be further extended to the left.
}

\end{document}